\documentclass[a4paper,12pt]{article}
\usepackage{amssymb}

\newtheorem{defn}{Definition}[section]
\newtheorem{theorem}[defn]{Theorem}
\newtheorem{lemma}[defn]{Lemma}
\newtheorem{corollary}[defn]{Corollary}
\newenvironment{proof}
{\noindent{\it Proof.}}{\hfill $\Box$\par\vspace{2.5mm}}

\newcommand{\zero}{\mbox{$\bigcirc$\hskip -2.9 mm {\small{0}}}}
\newcommand{\one}{\mbox{$\bigcirc$\hskip -2.8 mm {\small{1}}}}

\begin{document}

\begin{center}
{\Large
TROPICAL NEVANLINNA THEORY
\vskip 2mm
 AND 
 \vskip 4mm
 ULTRA-DISCRETE EQUATIONS}
\vskip 7mm
{\large
R. G. Halburd and N. J. Southall}
\vskip 5mm
Department of Mathematical Sciences, Loughborough University,
Loughborough, Leicestershire, LE11~3TU, UK
\vskip 5mm
{\tt R.G.Halburd@lboro.ac.uk}

\vskip 9mm

\begin{abstract}
A tropical version of Nevanlinna theory is described in which the role of meromorphic functions is played by continuous piecewise linear functions of a real variable whose one-sided derivatives are integers at every point.
These functions are naturally defined on the max-plus (or tropical) semi-ring.  Analogues of the Nevanlinna characteristic, proximity and counting functions are defined and versions of Nevanlinna's first main theorem, the lemma on the logarithmic derivative and Clunie's lemma are proved.

As well as providing another example of a tropical or dequantized analogue of an important area of complex analysis, this theory has applications to so-called ultra-discrete equations.  Preliminary results are presented suggesting that the existence of  finite-order max-plus meromorphic solutions can be considered to be an ultra-discrete analogue of the
Painlev\'e property.

\end{abstract}

\end{center}

\section{Introduction}
The max-plus (or tropical) semi-ring is the set
$\mathbb{R}\cup\{-\infty\}$
with addition and multiplication defined by
$x\oplus y:=\max(x,y)$ and 
$x\otimes y:=x+y$.  The additive and multiplicative identities are
$\zero=-\infty$ and $\one=0$ respectively. 
This structure fails to be a ring because not all elements have additive inverses.  In particular,
the equation $x\oplus 2=1$ has no solution.

The max-plus semi-ring
first arose in Kleene's 1956 paper on nerve sets and automata \cite{kleene:56}.
Recently the so-called ``tropical approach'' to mathematics \cite{mikhalkin:06,richterst:05,speyers:04} has attracted much attention from researchers in several fields including combinatorics, optimization, mathematical physics and algebraic geometry.   Although there is no subtraction in the tropical semi-ring, it is possible to construct polynomials and rational functions.  Surprisingly, many results from classical algebraic geometry have tropical analogues \cite{viro:01}.  

Tropical algebraic geometry has been useful in the enumeration of real and complex algebraic curves
and in the computation of Gromov-Witten invariants \cite{gathmannm:05,kontsevichm:94}
 of toric surfaces.  The approach of Kontsevich and Mikhalkin \cite{Mikhalkin:05} is to reduce the enumeration of algebraic curves to a count of tropical curves.  In this point of view, tropical geometry describes extreme degenerations of complex structures.

The operations $\oplus$ and $\otimes$ often arise from certain limits of expressions involving the usual operations $+$ and $\times$.  Suppose that $X$, $Y$ and $Z$ are real numbers and let $x=\exp(X/\epsilon)$, $y=\exp(Y/\epsilon)$ and $z=\exp(Z/\epsilon)$.  If $x$, $y$, $z$ satisfy the equation $z=xy$, then $X$, $Y$, $Z$ satisfy the equation $Z=X+Y=X\otimes Y$.  On the other hand, if $z=x+y$, then in the limit in which $\epsilon$ tends to zero from above and $X$ and $Y$ are held fixed, we have $Z=\max(X,Y)=X\oplus Y$,
where the identity
$
\lim_{\epsilon\to 0^+}\epsilon\log\left\{
\exp(A/\epsilon)+\exp(B/\epsilon)
\right\}=\max(A,B)
$
has been used.
This process is often referred to as {\em dequantization} or {\em ultra-discretization}, since this limit can be viewed in certain quantum mechanical models as the limit in which Planck's constant tends to zero (in the imaginary direction.)

Recent work suggests that there is a deep connection between integrable cellular automata and tropical geometry.
The first integrable ultra-discrete equations were introduced in \cite{takahashis:90,takahashim:95} and are related to box and ball systems.  
Such equations are naturally expressed in terms of the max-plus semi-ring and arise through the ultra-discretization of known integrable discrete equations
 \cite{tokihirotms:96,matsukidairasttt:97}.  There has been particular interest in the ultra-discrete Painlev\'e equations \cite{takahashitgor:97,ramanitgo:98,grammaticosortt:97,joshino:04}, which are tropical versions of the discrete Painlev\'e equations.  Joshi and Lafortune \cite{joshil:05} have described an analogue of singularity confinement for ultra-discrete equations.

Nevanlinna theory studies the value distribution of meromorphic functions.
Given a meromorphic function $f$, the Nevanlinna characteristic $T(r,f)$ is a non-negative non-decreasing function of $r>0$.  The Nevanlinna characteristic is a measure of the ``affinity'' of $f$ for infinity on the disc $|z|\le r$.  It is the sum of two terms: $m(r,f)$, which is large when $f$ is large on average on the circle $|z|=r$, and $N(r,f)$, which is large when $f$ has many poles in the disc $|z|<r$.  The behaviour of $T(r,f)$, $m(r,f)$ and $N(r,f)$ as $r$ tends to infinity encodes a great deal of information about $f$.  An important class of meromorphic functions are those for which the Nevanlinna characteristic is bounded by $r^\sigma$, for some $\sigma$.  Such functions are said to be of finite order.  We will define natural analogues of all of these concepts for the max-plus semi-ring and prove analogues of 
Nevanlinna's first main theorem, the lemma on the logarithmic derivative and Clunie's lemma.  The definition of a max-plus meromorphic function will be motivated in section 2.  We will define what it means for a continuous piecewise-linear function to be max-plus meromorphic on $\mathbb{R}$ and on $\mathbb{R}\cup\{-\infty\}$.  Nevanlinna theory can be defined for functions that are max-plus meromorphic on either of these domains.  In the present paper we will concentrate on functions max-plus meromorphic on $\mathbb{R}$ as this appears to be the form most applicable to the theory of ultra-discrete equations.

The lemma on the logarithmic derivative leads to Nevanlinna's second main theorem, which is a very powerful generalization of Picard's Theorem.  Difference analogues of these theorems were presented in \cite{halburdk:05jmaa,chiangf:06,halburdk:06fenn}.
We prove a max-plus version of the lemma on the logarithmic difference and then derive an analogue of Clunie's lemma, which is important in applications to ultra-discrete equations.   The basic set up of tropical Nevanlinna theory will be given in section 3.  Section 4 will concentrate on applications to ultra-discrete equations.

In \cite{ablowitzhh:00}, Ablowitz, Halburd and Herbst suggested that a difference equation should be considered to be of Painlev\'e type if it possesses sufficiently many finite-order meromorphic solutions.
Some apparently integrable discrete analogues of the (differential) Painlev\'e equations have been known for some time \cite{ramanigh:91}.  Let $R(z,y)$ be a rational function of  $y$
with coefficients meromorphic in $z$.
If an equation of the form 
\begin{equation}
y(z+1)+y(z-1)=R(z,y(z))
\label{classicalsum}
\end{equation}
has an admissible finite-order meromorphic solution then either $y$ satisfies a (linearizable) difference Riccati equation or equation (\ref{classicalsum}) is essentially one of the known difference Painlev\'e equations \cite{halburdk:07}.  

We will present preliminary results suggesting that 
the existence of sufficiently many finite-order max-plus meromorphic solutions is a natural
analogue of the Painlev\'e property for ultra-discrete equations
(although the existence of just one admissible solution might not be enough to get an analogue the the theorem mentioned above on the classification of (\ref{classicalsum}).)
Numerical evidence suggests that this criterion is sensitive to
the form of the coefficient functions in ultra-discrete equations.  Connections with the conditions  obtained by Joshi and Lafortune based on their analogue of singularity confinement
\cite{joshil:05} will be described.


\section{Max-plus meromorphic functions}
A natural definition of a max-plus rational function is a function of the form
$$
R(x)=\{a_0\oplus a_1\otimes x\oplus\cdots\oplus a_p\otimes x^{\otimes p}\}
\oslash
\{b_0\oplus b_1\otimes x\oplus\cdots\oplus b_q\otimes x^{\otimes q}\},
$$
where $x\oslash y:=x-y$, $x^{\otimes n}:=nx$ and $p$ and $q$ are non-negative integers.
Any max-plus rational function is a continuous piecewise linear function with only a finite number of distinct linear segments, each with integer slope.  This motivates the following definition.
\begin{defn}
A continuous piecewise linear function $f:\mathbb{R}\to\mathbb{R}$  is said to be {\em max-plus meromorphic on} 
$\mathbb{R}$ if 
both one-sided derivatives are integers at each point $x\in\mathbb{R}$.  
\end{defn}

For each $x\in\mathbb{R}$ let 
$\omega_f(x)=\lim_{\epsilon\to 0^+}
\left\{
f'(x+\epsilon)-f'(x-\epsilon)
\right\}$.  If $\omega_f(x)>0$ then $x$ is called a {\em root} of $f$ with multiplicity $\omega_f(x)$.
If $\omega_f(x)<0$ then $x$ is called a {\em pole} of $f$ with multiplicity $-\omega_f(x)$.  

Consider the function $f(x)=x\oplus a=\max\{x,a\}$, which has a root at $a$.  From the point of view of algebra and factorization, it is perhaps more natural to think of $a$ as the negative of the root.  However, there is no subtraction in the max-plus semi-ring and the definition given above is the natural one from the point of view of geometry.

If $f$ is max-plus meromorphic on $\mathbb{R}$ and $f'(x)=m$ for all $x<x_0$, for some constants
$m\in \mathbb{Z}$ and $x_0\in\mathbb{R}$ then we say that $f$ is max-plus meromorphic on
$\mathbb{R}\cup\{-\infty\}$.  The point $-\infty$ is called an {\em ordinary point} if $m\ge 0$, a {root} of order $m$, if $m>0$, and a pole of order $-m$ if $m<0$.

Nevanlinna theory can be developed for functions max-plus meromorphic on $\mathbb{R}$ and
for functions max-plus meromorphic on $\mathbb{R}\cup\{\infty\}$.  Although the latter is perhaps more natural from the point of view of generalizing certain series representations (power series, Hadamard decompositions, etc.) and for making the most precise link with the usual theory through de-quantization/ultra-discretization, the former appears to be more relevant to our main application --- ultra-discrete equations.  One heuristic reason for this is the following.  Most of the ultra-discrete Painlev\'e equations in the literature arise from the ultra-dsicretization of $q$-difference equations.  The ultra-discrete limit is usually applied directly to the $q$-discrete equation without first converting the equation to shift form through an exponential change of independent variable.  Although shift Painlev\'e difference equations are believed to have many finite-order meromorphic solutions, the solutions of $q$-difference equations usually have (fixed) non-pole singularities at the origin.  In the ultra-discrete limit, the origin is mapped to $-\infty$.

\begin{lemma}
\label{l-rational}
If a  function is max-plus meromorphic on $\mathbb{R}$ then it is max-plus rational if and only if it has a finite number of roots and poles.
\end{lemma}

Let $f$ and $g$ be max-plus meromorphic functions, then  $f\oplus g$, $f\otimes g$ and $h:=f\circ g$
are max-plus meromorphic.  Furthermore, if $g\not\equiv -\infty$, then $f\oslash g$ is also
max-plus meromorphic.
Finally, let $R(x,y)$ be a max-plus rational function in $y$ with coefficients that are max-plus meromorphic in $x$.
That is, $R$ has the form
\begin{eqnarray}
R(x,y)
&=&
({a_0(x)\oplus a_1(x)\otimes y\oplus \cdots\oplus a_p(x)\otimes y^{\otimes p}})
\nonumber\\
&&\oslash
({b_0(x)\oplus b_1(x)\otimes y\oplus \cdots\oplus b_q(x)\otimes y^{\otimes q}}),
\label{Rform}
\end{eqnarray}
where the $a_i$'s and $b_j$'s are max-plus meromorphic functions.  If $y$ is max-plus meromorphic
then so is $R(x,y(x))$.

\section{Tropical Nevanlinna Theory}
The starting point for Nevanlinna Theory is the Poisson-Jensen Formula, which expresses the value of the logarithm of a meromorphic function $f$ at some point $z$ in the disc $|z|<r$ in terms of the zeros and poles of $f$ in the disc and the average of a certain expression involving $f$ on the boundary of the disc (see, e.g., Hayman \cite{hayman:64}.)
The following is a natural analogue for max-plus meromorphic functions.
\begin{lemma}
Suppose that $f$ is a max-plus meromorphic function on $[-r,r]$, for some $r>0$ and denote the roots of $f$ in this interval by $a_\mu$, $\mu = 1,\ldots, M$, and the poles by $b_\nu$, $\nu=1,\ldots,N$, where roots and poles are listed according to their multiplicities.
Then for any $x\in(-r,r)$ we have the
{\em max-plus Poisson-Jensen Formula}
\begin{eqnarray}
&&
\qquad\quad
f(x)=\frac12\left\{
f(r)+f(-r)
\right\}
+
\frac{x}{2r}\left\{
f(r)-f(-r)
\right\}
\nonumber
\\
&&
-\frac1{2r}\sum_{\mu=1}^M\left\{
r^2-|a_\mu-x|r-a_\mu x
\right\}
+
\frac1{2r}\sum_{\nu=1}^N\left\{
r^2-|b_\nu-x|r-b_\nu x
\right\}.\qquad
\label{mPJ}
\end{eqnarray}
In particular, the case $x=0$ gives the {\em max-plus Jensen Formula}
\begin{equation}
f(0)=\frac{f(r)+f(-r)}{2}-\frac12\sum_{\mu=1}^M(r-|a_\mu|)+\frac12\sum_{\nu=1}^N(r-|b_\nu|).
\label{mJ}
\end{equation}
\end{lemma}

\setlength{\unitlength}{1pt}

\begin{figure}[ht]
\caption{Notation used in the proof of the max-plus Poisson-Jensen formula
\label{pl-pic}
}
\begin{picture}(400,200)
\put(0,110){\vector(1,0){360}}
\put(180,107.3){\small$\circ$}
\put(183,103){\makebox(0,0){\small$0$}}
\put(10,108){\line(0,1){4}}
\put(36,34){\makebox(0,0){\small$m_{-p-1}$}}
\put(10,104){\makebox(0,0){\small$-r$}}
\put(10,30){\line(1,1){20}}
\put(30,104){\makebox(0,0){\small$c_{-p}$}}
\put(30,108){\line(0,1){4}}
\put(30,50){\line(1,0){30}}
\put(45,56){\makebox(0,0){\small$m_{-p}$}}
\put(60,50){\line(1,3){30}}
\put(60,108){\line(0,1){4}}
\put(60,104){\makebox(0,0){\small$c_{-p+1}$}}
\put(85,70){\makebox(0,0){\small$m_{-p+1}$}}
\put(95,120){\makebox(0,0){$\cdots$}}
\put(110,120){\line(1,-2){20}}
\put(110,95){\makebox(0,0){\small$m_{-2}$}}
\put(130,80){\line(1,0){20}}
\put(140,85){\makebox(0,0){\small$m_{-1}$}}
\put(130,104){\makebox(0,0){\small$c_{-1}$}}
\put(130,108){\line(0,1){4}}
\put(150,80){\line(1,-1){20}}
\put(150,104){\makebox(0,0){\small$x$}}
\put(150,108){\line(0,1){4}}
\put(169,72){\makebox(0,0){\small$m_1$}}
\put(170,60){\line(1,0){40}}
\put(190,65){\makebox(0,0){\small$m_2$}}
\put(170,104){\makebox(0,0){\small$c_1$}}
\put(170,108){\line(0,1){4}}
\put(210,60){\line(1,3){35}}
\put(210,104){\makebox(0,0){\small$c_2$}}
\put(225,133){\makebox(0,0){\small$m_3$}}
\put(210,108){\line(0,1){4}}
\put(245,165){\line(1,1){15}}
\put(247,180){\makebox(0,0){\small$m_4$}}
\put(245,104){\makebox(0,0){\small$c_3$}}
\put(245,108){\line(0,1){4}}
\put(267,170){\makebox(0,0){$\cdots$}}
\put(275,185){\line(1,-2){15}}
\put(295,175){\makebox(0,0){\small$m_{q-1}$}}
\put(290,155){\line(1,0){20}}
\put(295,104){\makebox(0,0){\small$c_{q-1}$}}
\put(300,148){\makebox(0,0){\small$m_q$}}
\put(295,108){\line(0,1){4}}
\put(310,155){\line(1,-1){40}}
\put(340,140){\makebox(0,0){\small$m_{q+1}$}}
\put(315,104){\makebox(0,0){\small$c_q$}}
\put(315,108){\line(0,1){4}}
\put(350,104){\makebox(0,0){\small$r$}}
\put(350,108){\line(0,1){4}}
\end{picture}
\vskip -16mm

\end{figure}
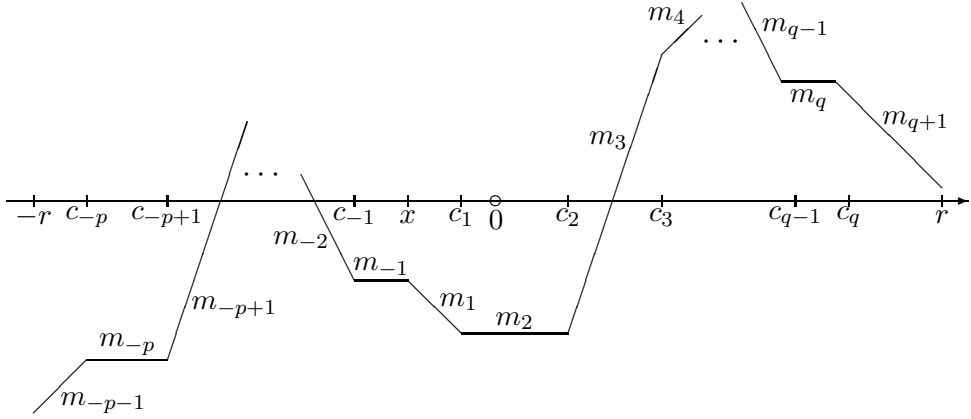

\vskip 10 mm

\begin{proof}
Define a finite increasing sequence of points $(c_k)_{k=-p,\ldots,q}$ as follows.
Let $c_0=x$ and let the other $c_k$'s denote the points in $\gamma\in(-r,r)$ at which $f'(\gamma)$
does not exist.  We denote by $m_k$ the slopes of the line segments in the graph of $f$.  Specifically,
for $k=-p,\ldots,0$ we set $m_{k-1}=\lim_{x\to c_k^-}f(x)$ and for $k=0,\ldots,q$ we
set $m_{k+1}=\lim_{x\to c_k^+}f(x)$ (see figure \ref{pl-pic}.)
It follows that
\begin{eqnarray}
f(r)-f(x)
&=&m_1(c_1-x)+m_2(c_2-c_1)+\cdots+m_q(c_q-c_{q-1})+m_{q+1}(r-c_q)\nonumber\\
&=&
-m_1x+m_{q+1}r+c_1(m_1-m_2)+\cdots+c_q(m_q-m_{q+1})\nonumber\\
&=&
m_1(r-x)-\sum_{j=1}^q(m_j-m_{j+1})(r-c_j),
\label{rminusx}
\end{eqnarray}
where we have used the fact that $m_{q+1}=m_1-\sum_{j=1}^q (m_j-m_{j+1})$.  Similarly we find that
\begin{equation}
\label{xminusr}
f(x)-f(-r)=
m_{-1}(r+x)+\sum_{i=1}^p (m_{-i-1}-m_{-i})(r+c_{-i}).
\end{equation}
Multiplying equations (\ref{rminusx}) and (\ref{xminusr}) by $(r+x)$ and $(r-x)$  respectively and then taking the difference gives
\begin{eqnarray*}
2rf(x)
&=&
r\left\{f(r)+f(-r)\right\}
+x\left\{f(r)-f(-r)\right\}
+(m_{-1}-m_1)(r^2-x^2)\\
&&
+\sum_{i=1}^p(m_{-i-1}-m_{-i})
(r^2-[x-c_{-i}]r-c_{-i}x)\\
&&
+\sum_{j=1}^q(m_j-m_{j+1})(r^2-[c_j-x]r-c_jx).
\end{eqnarray*}
\vskip - 7mm
\end{proof}

For any $x\in\mathbb{R}\cup\{-\infty\}$ we define $x^+:=\max\{x,0\}$.  For any function $f:{\mathbb{R}}\to{\mathbb{R}}$, denote by $f^+$ the function given by
$
f^+(x):=\max(f(x),0).
$
The {\em max-plus proximity function} is then defined to be
$$
m(r,f)=\frac{f^+(r)+f^+(-r)}{2}.
$$
Some useful identities involving the max-plus proximity function are
$m(r,f)\le (|f(r)|+|f(-r)|)/2$,
$m(r,f\oplus g)=m(r,\max\{f(x),g(x)\})\le m(r,f)+m(r,g)$
and
$m(r,f\otimes g)=m(r,f+g)\le m(r,f)+m(r,g)$.

The {\em max-plus counting function}, $n(r,f)$, gives the number of poles of $f$ in the interval $(-r,r)$, counting multiplicities.  The {\em integrated max-plus counting function} is defined to be
$$
N(r,f):=\frac12\int_0^r n(t,f)\,dt=\frac12\sum_{\nu=1}^N(r-|b_\nu|).
$$
It follows from the max-plus Jensen formula (\ref{mJ}) that the {\em max-plus characteristic function},
$T(r,f):=m(r,f)+N(r,f)$,
satisfies
\begin{equation}
\label{almost-first}
T(r,f)-T(r,-f)=f(0).
\end{equation}
Equation (\ref{almost-first}) is a weak analogue of Nevanlinna's first main theorem.
Note that $-f=\one\,\oslash f$.  
Let $L:=\inf\{f(b)\,:\, b\mbox{ is a pole of }f\}$.  If $L>-\infty$ then for all $a<L$, we have an analogue of the first main theorem, namely
$T(r,\one\,\oslash(f\,\oplus a))=T(r,f)+O(1)$.
The max-plus Nevanlinna characteristic satisfies a number of useful estimates such as $T(r,f\oplus g)=T(r,\max(f,g))\le T(r,f)+T(r,g)$ and
$T(r,f\otimes g)=T(r,f+g)\le T(r,f)+T(r,g)$.

\begin{lemma}
$T(r,f)$ is a continuous non-decreasing piecewise linear function of $r$.
\end{lemma}

\begin{proof}
Choose $r>0$ such that $f$ does not have a root or a pole at either $x=r$ or $x=-r$.
Differentiating the max-plus Jensen formula (\ref{mJ}) gives
\begin{equation}
\label{Td}
f'(r)-f'(-r)=n(r,-f)-n(r,f).
\end{equation}
If $f(-r)<0$ and $f(r)<0$, then
$$
\frac{dT(r,f)}{dr}=\frac 12 n(r,f)\ge 0.
$$
Also, if $f(-r)\ge0$ and $f(r)\ge0$, then
$$
\frac{dT(r,f)}{dr}=\frac 12 \left\{f'(r)-f'(-r)+n(r,f)\right\}=\frac12 n(r,-f)\ge 0,
$$
where we have used equation (\ref{Td}).
Next we consider the case in which $f(-r)<0$ and $f(r)\ge0$.  In this case there must be a sub-interval of $(-r,r)$ on which the graph of $f$ has strictly positive slope.  Therefore, the slope of the graph of $f$ at $x=r$ is strictly greater than $-n(r,f)$.  It follows that
$$
\frac{dT(r,f)}{dr}=\frac 12 \left\{f'(r)+n(r,f)\right\}>0.
$$
Finally we consider the case in which $f(-r)\ge 0$ and $f(r)<0$.  Similar reasoning to that in the previous case shows that $f'(-r)<n(r,f)$.  Hence
$$
\frac{dT(r,f)}{dr}=\frac 12 \left\{-f'(-r)+n(r,f)\right\}\ge 0.
$$
\end{proof}

\begin{lemma}
\label{nlemma}
For any $k>1$, the counting function satisfies
$$
n(r,f)\le \frac 2{(k-1)r}N(kr,f).
$$
\end{lemma}
\begin{proof}
$$
(k-1)r\,n(r,f)=n(r,f)\int_r^{kr}dt\le\int_r^{kr}n(t,f)\,dt
\le
\int_0^{kr}n(t,f)\,dt=2N(kr,f).
$$
\end{proof}

\begin{theorem}
Let $f$ be a max-plus meromorphic function.  Then $T(r,f)=O(r)$ if and only if $f$ is a max-plus rational function.
\end{theorem}

\begin{proof}
If $f$ is a max-plus rational function then there exists $R>0$ such that
$f(r)=A_1r+A_2$, $f(-r)=A_3r+A_4$ and $n(r,f)=A_5$, for certain constants $A_1,\ldots,A_5$ and for all $r>R$.  The result follows immediately from the definitions of $N(r,f)$ and $T(r,f)$.

Next we assume that $f$ is a meromorphic function satisfying $T(r,f)\le Kr$, for some $K$ and all sufficiently large $r$.  
From Lemma \ref{nlemma} we have, for any $k>1$,
$$
n(r,f)\le \frac 2{(k-1)r}N(kr,f)
\le \frac 2{(k-1)r}T(kr,f)\le
 \frac {2kK}{(k-1)}.
$$
Hence $f$ has a finite number of poles.  Similarly, using equation
(\ref{almost-first}), we find that $n(r,-f)$ is also bounded and so $f$ has a finite number of roots.  It follows from Lemma \ref{l-rational} that $f$ is rational.
\end{proof}
Other classes of max-plus meromorphic functions
have particular growth properties as measured by $T(r,f)$.
For example, for any non-constant periodic max-plus meromorphic function there exist positive constants $K_1$ and $K_2$ such that $K_1r^2\le T(r,f) \le K_2 r^2$, for sufficiently large $r$.
This is much more rigid than the complex analytic case.  Periodic functions can be infinite order in the complex plane.  This fact has important consequences for applications of Nevanlinna theory to difference equations.

\begin{lemma}
\label{borel}
(Generalized Borel Lemma \cite[Lemma
3.3.1]{cherryy:01})\\
Let $\xi(x),$ and $\phi(r)$ be positive, nondecreasing,
continuous functions defined for $e\leq x<\infty$ and $r_0\leq
r<\infty$, respectively, where $r_0$ is such that $T(r)\geq
e$ for all $r\geq r_0$, for some positive nondecreasing continuous function T.  Then
    $$
    T\left(r+\frac{\phi(r)}{\xi(T(r))}\right) \leq 2 T(r)
    $$
for all $r$ outside a
 (possibly empty) 
set $E$ such that, for all $R<\infty$,
    $$
    \int_{E\cap [r_0,R]} \frac{dr}{\phi(r)} \leq \frac{1}{\xi(e)}+ \frac{1}{\log
     2}\int_e^{T(R)}\frac{dx}{x\xi(x)}.
    $$
\end{lemma}
\noindent
In particular, this lemma implies the standard Borel Lemma (see, e.g., \cite{hayman:64}), which says that
\begin{equation}
\label{borelineq}
T\left(r+\frac{1}{T(r)}\right)\le 2T(r),
\end{equation}
outside an exceptional set of finite linear measure.

\begin{theorem}
\label{nNnoorder}
If $f$ is max-plus meromorphic then for any $\epsilon>0$,
$n(r,f)\le 4 r^{-1} N(r,f)^{1+\epsilon}$,
outside  an exceptional set $E$ of finite logarithmic measure. 
\end{theorem}

\begin{proof}  If $n(r,f)\equiv 0$, there is nothing to prove.  From Lemma \ref{nlemma} with 
$k=1+N(r,f)^{-\epsilon}$,
we have 
$$
n(r,f)\le \frac{2}{r}N(r,f)^\epsilon N\left(r+\frac r{N(r,f)^\epsilon},f\right).
$$
Now we apply Lemma \ref{borel} with $T(r)=N(r,f)$, $\phi(r)=r$ and $\xi(x)=x^\epsilon$, which shows that
$$
N\left(r+\frac r{N(r,f)^\epsilon},f\right)\le 2N(r,f),
$$
outside  an exceptional set $E$ satisfying
$$
 \int_{E\cap [r_0,R]} \frac{dr}{r} \leq \frac{1}{e^\epsilon}+ \frac{1}{\log
     2}\int_e^{N(R,f)}\frac{dx}{x^{1+\epsilon}}
     \le
     \left(1+\frac{1}{\epsilon\log 2}\right)e^{-\epsilon}.
$$
\end{proof}

\begin{defn}
A max-plus meromorphic function is said to be of finite order if
there exist positive numbers $\sigma$ and $r_0$ such that
$T(r,f)\le r^\sigma$, for all $r>r_0$.
\end{defn}

\begin{corollary}
\label{ncorollary}
Let $f$ be a finite-order max-plus meromorphic function.  Then for all $\delta<1$,
$n(r,f)\le r ^{-\delta} {N(r,f)}$,
outside  an exceptional set $E$ of finite logarithmic measure.
\end{corollary}

\begin{proof}
Now $N(r,f)\le T(r,f)\le r^{\sigma}$.
Choose $\epsilon<(1-\delta)/\sigma$.  Then for sufficiently large $r$,
$4N(r,f)^\epsilon < r^{1-\delta}$.  Now apply Theorem \ref{nNnoorder}.
\end{proof}

The finite-order condition is important in Corollary \ref{ncorollary}.  Consider the infinite-order function $f$ such that $f(x)=0$ for all $x<0$, $f$ has no roots and its only poles occur at each non-negative integer $n$ with multiplicity $2^n$.  In this case $N(r,f)=O(n(r,f))$.

The following lemma and the subsequent theorem constitute natural max-plus difference analogues of the lemma on the logarithmic derivative.

\begin{lemma}
Let $f$ be a  max-plus meromorphic function. Then for any $\epsilon>0$,
$$
m(r,f(x+c)\oslash f(x))\le \frac{2^{1+\epsilon}\cdot 14 |c|}{r}
\left\{
T(r+|c|,f)^{1+\epsilon}+o(T(r+|c|,f)
\right\},
$$
outside an exceptional set of finite logarithmic measure.
\end{lemma}

\begin{proof}
For any $\rho>r+|c|$ and $x\in [-r,r]$, the max-plus Poisson-Jensen formula gives
\begin{eqnarray*}
f(x+c)-f(x)
&=&
\frac{c}{2\rho}\left\{ f(\rho)-f(-\rho)\right\}\\
&&
+
\frac 1{2\rho}\sum_{\mu} \left\{ \left(|a_\mu-x-c|-|a_\mu-x|\right)\rho+ca_\mu\right\}\\
&&
-
\frac 1{2\rho}\sum_{\nu} \left\{ \left(|b_\nu-x-c|-|b_\nu-x|\right)\rho+cb_\nu\right\}.
\end{eqnarray*}
Therefore
\begin{eqnarray*}
m(r,f(x+c)-f(x))
\le  \left(\frac{c}{2\rho}\left\{ f(\rho)-f(-\rho)\right\}\right)^+\\
\qquad+
\sum_{\mu}m\left(r,
\frac 1{2\rho} \left\{ \left(|a_\mu-x-c|-|a_\mu-x|\right)\rho+ca_\mu\right\}\right)\\
\qquad+
\sum_{\nu} m\left(r,-
\frac 1{2\rho} \left\{ \left(|b_\nu-x-c|-|b_\nu-x|\right)\rho+cb_\nu\right\}\right).
\end{eqnarray*}
Now
\begin{eqnarray*}
 \left(\frac{c}{2\rho}\left\{ f(\rho)-f(-\rho)\right\}\right)^+
\le
\left|\frac{c}{2\rho}\left\{ f(\rho)-f(-\rho)\right\}\right|\\
\qquad\le
\frac{|c|}{2\rho}\left\{ f^+(\rho)+f^+(-\rho)+(-f)^+(\rho)+(-f)^+(-\rho)\right\}\\
\qquad=\frac{|c|}{\rho}\left\{m(\rho,f)+m(\rho,-f)\right\}
.
\end{eqnarray*}
Also
\begin{eqnarray*}
&&m\left(r,
\frac 1{2\rho} \left\{ \left(|a_\mu-x-c|-|a_\mu-x|\right)\rho+ca_\mu\right\}\right)\\
&\le&
\frac12\left(\left\{
\big|
|a_\mu-r-c|-|a_\mu-r|
\big|
+
\big|
|a_\mu+r-c|-|a_\mu+r|
\big|
\right\}
+
\frac{|ca_\mu |}{\rho}
\right)
\le
\frac32|c|,
\end{eqnarray*}
since $|a_\mu|<\rho$.
From the above estimates and Theorem \ref{nNnoorder},
 for any $\epsilon>0$,
\begin{eqnarray*}
&&m(r,f(x+c)-f(x))\\
&\le&
|c|\left\{\frac1\rho\left(m(\rho,f)+m(\rho,-f)\right)+\frac32 \left(n(\rho,f)+n(\rho,-f)\right)\right\}
\\
&\le&
\frac{|c|}{\rho}\left\{
T(\rho,f)+T(\rho,-f)+6T(\rho,f)^{1+\epsilon}+6T(\rho,-f)^{1+\epsilon}
\right\}
\\
&\le&
\frac{7|c|}{\rho}
\left\{
T(\rho,f)^{1+\epsilon}+T(\rho,-f)^{1+\epsilon}
\right\}\\
&\le&
\frac{14|c|}{\rho}\left\{T(\rho,f)^{1+\epsilon}+o\left(T(\rho,f)\right)\right\},
\end{eqnarray*}
outside  an exceptional set of finite logarithmic measure.  Choosing
$\rho=r+|c|+1/T(r+|c|,f)$ and using the Borel Lemma \ref{borel} with $r$ replaced by $r+|c|$, we obtain
$T(\rho,f)\le 2T(r+|c|,f)$,
outside a set of finite linear measure.  
\end{proof}

\begin{lemma}\label{technical}\cite[Lemma~2.1]{halburdk:07}
Let $T:\mathbb{R}^+\to\mathbb{R}^+$ be a non-decreasing continuous
function, $s>0$, $\alpha<1$, and let $F=\{r\in\mathbb{R}^{+}\,:\,    T(r) \leq \alpha T(r+s)\}$.
If the logarithmic measure of is $F$ infinite, that is,
$\int_F\frac{dt}{t}=\infty$, then
    $
    \limsup_{r\to\infty}{\log T(r)}/{\log r}=\infty.
    $
\end{lemma}

\begin{theorem}
\label{loglemma}
Given $\delta<1$, any finite-order max-plus meromorphic function $f$ satisfies
$m(r,f(x+c)\oslash f(x)) = O\left(r^{-\delta}T(r,f)\right)$,
outside  an exceptional set of finite logarithmic measure.
\end{theorem}

\begin{proof}
Since $f$ has finite order, Lemma \ref{technical} implies that $T(r+|c|,f)\le 2T(r,f)$ outside  an exceptional set of finite logarithmic measure.
Also, there exist positive constants $\sigma$ and $r_0$ such that
$T(r,f)\le r^\sigma$ for all $r>r_0$.  Choose $\epsilon=(1-\delta)/\sigma$.
Then $T(r,f)^\epsilon/r\le r^{-\delta}$.
\end{proof}

\section{Applications to ultra-discrete equations}
The main application that we have in mind for tropical Nevanlinna theory is as a measure of the complexity of solutions of ultra-discrete equations.  In particular, the aim is to use ideas from tropical Nevanlinna theory to classify equations that are natural ultra-discrete analogues of the Painlev\'e equations.  
Many such equations have been considered in the literature recently \cite{grammaticosortt:97,ramanitgo:98,joshino:04,joshil:05,joshil:06}.  Most of these equations have been obtained directly as ultra-discretizations of known discrete Painlev\'e equations.
Examples of such equations include
\begin{eqnarray}
y_{n+1}+y_{n-1}&=&\max\{y_n+n,0\}-y_n,\label{udpone}\\
y_{n+1}+y_{n-1}&=&a+\max\{y_n,n\}-\max\{y_n+n,0\}-y_n,\label{udptwo}\\
y_{n+1}+y_{n-1}&=&\max\{ n+a,y_n\}+\max\{n-a,y_n\}\nonumber\\
&& \ \ \ -\max\{y_n+n+b,0\}-\max\{y_n+n-b,0\},\label{udpthree}
\end{eqnarray}
where $a$ and $b$ are constants.

Conventionally, only solutions of ultra-discrete equations that are functions from $\mathbb{Z}$ to itself are considered.  However, equations such as (\ref{udpone}--\ref{udpthree}) can be re-interpreted as
equations for a continuous piecewise linear real function $y$ of a real variable $x$.  In particular,
instead of equation (\ref{udpone}), we consider the ``extended'' equation
\begin{equation}
\label{new}
y(x+1)+y(x-1)=\max\{y(x)+x,0\}-y(x),\qquad x\in\mathbb{R}.
\end{equation}
It now makes sense to ask about the existence and general properties of max-plus meromorphic solutions of
equations such as (\ref{new}).  Based on analogous considerations of (genuine) meromorphic solutions of difference equations in the complex domain in \cite{ablowitzhh:00,grammaticostrt:01,halburdk:07}, it is natural to begin by considering ultra-discrete equations admitting finite-order max-plus meromorphic solutions.
We will present evidence that this property can be thought of as an ultra-discrete analogue of the Painlev\'e property.  The Painlev\'e property is property is closely associated with the integrability of differential equations.  A more thorough study of this characterisation for ultra-discrete equations will appear in a future work
\cite{nev-ultra}.

We begin by addressing some simple questions on the existence of max-plus meromorphic solutions.
\begin{lemma}
\label{simplelemma}
The equation
\begin{equation}
\label{simplefirst}
y(x+1)=y(x)^{\otimes n}=ny(x)
\end{equation}
admits a non-constant max-plus meromorphic solution on $\mathbb{R}$ if and only if
$n=\pm 1$.
\end{lemma}

\begin{proof}
If $n=0$ then $y\equiv 0$ is the only solution.  Recall that any periodic max-plus meromorphic
function is of finite order.
If $n=1$ then $y$ is any max-plus meromorphic period one function.  If $w$ is any period two max-plus meromorphic function, then $y(x):=w(x+1)-w(x)$ is a max-plus meromorphic solution of equation (\ref{simplefirst}) with $n=-1$.

If $y$ is non-constant then $\exists x_0\in\mathbb{R}$ such that $y'$ exists and is a non-zero integer
$m$ at $x_0$.
It follows from equation (\ref{simplefirst}) that for all $\nu\in \mathbb{Z}$,
$y'(x_0-\nu)=m/n^{\nu}$.  Therefore if $\nu\ne \pm 1$ then for sufficiently large $\nu$,
$0<|y'(x_0-\nu)|<1$, and hence the slope is not an integer.
\end{proof}

Note that the max-plus Nevanlinna characteristic can be defined for arbitrary continuous piecewise linear
 functions (not necessarily with integer slopes) if we allow the counting function $n(r,f)$ to count poles of non-integer multiplicites (i.e., the differences in slopes).  This point of view will be explored further in 
 \cite{nev-ultra}.  For now we remark that allowing for non-integer multiplicities, the extra condition of finite-order needs to be added to the assumptions in lemma \ref{simplelemma} in order to reach the same conclusion.
 
 Apart from the analogue of Clunie's lemma \ref{clunie} below, we shall restrict our attention to ultra-discrete equations
 of the form 
 \begin{equation}
 \label{second-order}
 y(x+1)\otimes y(x-1)=R(x,y(x)),
 \end{equation}
where $R$ is max-plus rational in $x$ and $y$.  We remark that all such equations admit infinitely many max-plus meromorphic solutions.  To see this, choose $y(0)$ and $y(1)$ to be any real numbers and calculate $y(2):=R(1,y(1))-y(0)$.  Now define $y$ on $(0,1)\cup(1,2)$ such that $y$ is a continuous piecewise-linear function on $[0,2]$ with integer slopes wherever $y'$ is defined.  Then the equation itself extends $y$ uniquely to a max-plus meromorphic solution on
 $\mathbb{R}$.

We will show that large classes of equations of the form (\ref{second-order})
admit infinite-order solutions.  In the simplest cases, this can be achieved by showing that there is a sequence of integers $(\nu_n)$ such that $|\nu_n|\to \infty$ and
$y(x_0+\nu_n)\ge C^{\nu_n}$ for some $C>1$.

\begin{lemma}
Let $y\not\equiv 0$ be a max-plus meromorphic solution of
\begin{equation}
\label{nomax}
y(x+1)\otimes y(x-1)=y(x)^{\otimes n},
\end{equation}
for some $n\in\mathbb{Z}$.  If $y$ is of finite order then $|n|\le 2$.
\end{lemma}

\begin{proof}
Let
$$
\lambda_\pm=\frac{n\pm\sqrt{n^2-4}}{2}.
$$
$\exists x_0\in\mathbb{R}$ such that $y(x_0)\ne 0$.  Therefore, for at least one choice of ``$+$'' or ``$-$'',
we have that
$y(x_0+1)\ne \lambda_\pm y(x_0)$.  Then for each $\nu\in\mathbb{Z}$,
\begin{equation}
\label{linearsoln}
y(x_0+\nu)=\alpha \lambda_+^\nu+\beta \lambda_-^\nu,
\end{equation}
where
$$
\alpha=\frac{y(x_0+1)-\lambda_-y(x_0)}{\lambda_+-\lambda_-}\quad\mbox{and}\quad
\beta=\frac{y(x_0+1)-\lambda_+y(x_0)}{\lambda_--\lambda_+}
$$
are not both zero. 

Now if $n>2$, then $\lambda_+>1$ and $\lambda^{-1}_->1$,
while if
$n<-2$ then $-\lambda_->1$ and $-\lambda^{-1}_+>1$.  Hence either $y(x_0+\nu)$ or
$y(x_0-\nu)$ grows exponentially as $\nu$ tends to infinity on the even integers.
So $T(r,y)$ is not bounded by a power of $r$.
\end{proof}

\begin{theorem}
Let $P(y)=\max\{a_0,a_1+y,\ldots,a_p+py\}$ and
$Q(y)=\max\{b_0,b_1+y,\ldots,b_q+qy\}$ be two max-plus polynomials   with no common roots
and neither $a_p$ nor $b_q$ is $-\infty$.  If $|p-q|>2$, then the equation
\begin{equation}
\label{general}
y(x+1)+y(x-1)=P(y(x))-Q(y(x))
\end{equation}
admits infinitely many infinite-order max-plus meromorphic solutions.
\end{theorem}

\begin{proof}
If $y$ is sufficiently large for all $x$ larger than some number $\xi$, then
equation (\ref{general}) reduces to
$$
y(x+1)+y(x-1)=(p-q)y(x)+a_p-b_q,
$$ 
for all $x>\xi$.  If $p-q>2$ then, given $\alpha>0$, there is a family of solutions such that any member  evaluated at an integer
$\nu>\xi$  has the form
$$
y(\nu)=\frac{b_q-a_p}{p-q-2}+\alpha\left(
\frac{(p-q)+\sqrt{(p-q)^2-4}}{2}
\right)^\nu.
$$
So $y$ and hence $T(r,y)$ grow exponentially.  If $p-q<-2$ then the same argument works with a minus sign in front of the square root.
\end{proof}

In  \cite{joshil:06}, Joshi and Lafortune consider the equation
$$
y_n+3y_n+y_{n-1}=\max\{x+K,0\},
$$
where $K$ is a constant, as an example of an ultra-discrete
equation that does not possess their singularity confinement property.  Analogously, we have the following.

\begin{lemma}
Let $K$ be a positive constant and let $y$ be a max-plus meromorphic solution of
\begin{equation}
\label{degreethree}
y(x+1)+3y(x)+y(x-1)=\max\{y(x)+K,0\}
\end{equation}
such that $y(0)>0$ and $y(1)<-K$.  Then $y$ has infinite order.
\end{lemma}

\begin{proof}
It is straightforward to show by induction that for all $n\ge 1$,
if $n$ is odd then $y(n+1)\ge -2y(n)>0$ and if $n$ is even then
$-y(n+1)\ge y(n)>0$.  Hence $y$ grows exponentially on $\mathbb{N}$.
\end{proof}

In \cite{joshil:05}, Joshi and Lafortune also considered the ultra-discrete equation
$$
X_{n-1}+X_n+X_{n+1}
=
\max\{X_n+\phi_n,0\}.
$$
and showed that the condition for singularity confinement is
$$
\phi_{n+5}-\phi_{n+3}-\phi_{n+2}+\phi_n=0.
$$
That is,
$$
\phi_{n}
=
\alpha+\beta n+\gamma (-1)^n
+\delta\cos\left(\frac{2\pi n}3\right)
+\omega\sin\left(\frac{2\pi n}3\right).
$$
We consider the analogous equation
\begin{equation}
\label{JLa}
y(x-1)+y(x)+y(x+1)
=
\max\{y(x)+\phi(x),0\}.
\end{equation}
The confinement criterion now becomes
$$
\phi(x)
=
\pi_2(x)+\pi_3(x)+Nx+C,
$$
where $\pi_2$ and $\pi_3$ are arbitrary periodic max-plus meromorphic functions of
period 2 and 3 respectively, and $N$ is an integer and $C$ is a real number.

Equation (\ref{JLa}) will be studied in greater detail in \cite{nev-ultra} however, here we note some important observations.  Analytically it an be shown that the solutions of equation (\ref{JLa}) are of finite-order if $\phi$ is a linear function.  Furthermore, numerical studies suggest that if $\phi$ is a periodic function of order 2 or 3 (or a sum of such functions) then the order of $y$ is finite.  If $\phi$ is chosen to be a max-plus meromorphic function of period 4 or 5 then $y$ appears to have infinite order.  However, when $\phi(x)$ is chosen to have the form $x+\psi(x)$, where $\psi$ is bounded, then numerical studies suggest that $y$ is finite order, regardless of the precise form of $\psi$.  However, in the cases studied, for sufficiently large $x$, the solutions of  equation (\ref{JLa}) are identical to (not merely asymptotic to) solutions of simpler ``integrable'' ultra-discrete equations.  This is quite unlike the complex analytic setting in which we have uniqueness of analytic continuation.

In \cite{halburdk:07}, the classification of difference equations admitting finite-order meromorphic solutions in the complex domain relied on estimating the relative distribution of the points at which the solution, $y$,  hits one of the finite singular values of the equation and the distribution of the poles of 
$y$.  The method used naturally led to a variant of the usual singularity confinement method.  An analogue of this part of the argument exists relating the distribution of the singular values where $y(x)=-\phi(x)$ to the poles of $y$, using method related to singularity confinement.  In order to deduce that non-confinement implies that the solution has infinite order, we need to show that there are ``many'' points at which the solution takes a singular value.  In the complex analytic  case \cite{halburdk:07}, this is guaranteed by using a difference version of Clunie's lemma and Nevanlinna's first main theorem.  Below we present an ultra-discrete version of Clunie's lemma, however, it is the absence of a strong max-plus version of Nevanlinna's first main theorem that prevents the same argument going through.  Indeed, when $\phi$ grows sufficiently fast it appears from numerical studies that any solution only hits a singular point a finite number of times.

In \cite{grammaticosrttc:07}, Grammaticos, Ramani, Tamizhmani, Tamizhmani and Carstea show that the equation
\begin{equation}
\label{confined}
y(x+1)=y(x-1)+|y(x)|
\end{equation}
possesses the ultra-discrete singularity confinement property.  The authors suggest that nevertheless this equation is not integrable.  From the max-plus Nevanlinna point of view, equation
(\ref{confined})
possesses infinite-order max-plus meromorphic solutions.  In particular, if $y(0)=y(1)=1$ then $(y(n))_{n\in\mathbb{N}}$ is the Fibonacci sequence, which grows exponentially.

The final result that we will present is an analogue of Clunie's lemma for ultra-discrete equations.
Let $\lambda=(\lambda_0,\lambda_1,\ldots,\lambda_m)$, where the $\lambda_j$s are non-negative integers, be a multi-index with respect to the shifts $(0,c_1,\ldots,c_m)\in\mathbb{R}^{m+1}$.  Let 
$$
f^{\otimes\lambda}(x)
:=
\lambda_0 f(x)+\lambda_1 f(x+c_1)+\cdots \lambda_m f(x+c_m).
$$
An expression of the form
$$
\sum_{\lambda\in\Lambda}a_{\lambda}(x)\otimes f^{\otimes\lambda}(x),
$$
where $\Lambda$ is a finite set of indices,
is called a max-plus polynomial in $f$ and its shifts.  We will say that the coefficients are small if
$T(r,a_\lambda)=o(T(r,f))$ outside a set of finite logarithmic measure.

The following is a natural analogue of Clunie's lemma.

\begin{theorem}
\label{clunie}
Let $P(x,f)$ and $Q(x,f)$ be max-plus polynomials in $f$ and its shifts with small coefficients.  If
$f$ is a finite-order max-plus meromorphic function satisfying
$f^{\otimes n}(x)P(x,f)=Q(x,f)$,
where the degree of $Q$ in $f$ and its shifts is less than or equal to $n$, then
for any $\delta<1$,
$$
m\left(r,P(x,f)\right)=O\left(r^{-\delta}
T(r,f)
\right)
+
o\left(
T(r,f)
\right),
$$
outside  an exceptional set of finite logarithmic measure.
\end{theorem}

\begin{proof}
Given $r>0$, let $S_+:=\{s\,:\, f(s)\ge 0\mbox{ and }|s|=r\}$ and
$S_-:=\{s\,:\, f(s)< 0\mbox{ and }|s|=r\}$.  Then
$$
m(r,P(x,f))=\frac12\left(
\sum_{s\in S_+}P(s, f)^++\sum_{s\in S_-}P(s, f)^+
\right).
$$
Let
$P(x,f)=\sum_{\lambda\in\Lambda_P}a_{\lambda}(x)\otimes f^{\otimes\lambda}(x)$
and
$Q(x,f)=\sum_{\lambda\in\Lambda_Q}b_{\lambda}(x)\otimes f^{\otimes\lambda}(x)$.
For any $x\in S_-$,
\begin{eqnarray*}
P(x,f)=\sum_{(\lambda_0,\ldots,\lambda_m)\in\Lambda_P}a_{\lambda}(x)\otimes
f^{\otimes\lambda_0}(x)\otimes f^{\otimes\lambda_1}(x+c_1)\otimes\cdots\otimes f^{\otimes\lambda_m}(x+c_m)
\\
\le
\max_{\stackrel{(\lambda_0,\ldots,\lambda_m)}{\in\Lambda_P}}\left\{
a_{\lambda}(x)+\lambda_1[f(x+c_1)-f(x)]
+\cdots+
\lambda_m[f(x+c_m)-f(x)]
\right\}.
\end{eqnarray*}
So using Lemma \ref{loglemma}, we see that
$$
\sum_{s\in S_-}P(s, f)^+=O\left(r^{-\delta}
T(r,f)
\right)
+
o\left(
T(r,f)
\right),
$$
outside  an exceptional set of finite logarithmic measure.  For $x\in S_+$, we note that
$P(x,f)=Q(x,f)-nf$ and the degree of $Q$ is at most $n$.  Hence
\begin{eqnarray*}
P(x,f)
\le
\max_{\stackrel{(\lambda_0,\ldots,\lambda_m)}{\in\Lambda_Q}}\left\{
b_{\lambda}(x)+\lambda_1[f(x+c_1)-f(x)]
+\cdots+
\lambda_m[f(x+c_m)-f(x)]
\right\}.
\end{eqnarray*}
So again using Lemma \ref{loglemma} we find that
$$
\sum_{s\in S_+}P(s, f)^+=O\left(r^{-\delta}
T(r,f)
\right),
$$
outside  an exceptional set of finite logarithmic measure.  
\end{proof}

\section{Conclusion}
We have introduced a max-plus version of Nevanlinna theory together with analogues of some of the key theorems that have been used in the classification of difference equations admitting finite-order max-plus meromorphic solutions in the complex domain.  The max-plus Nevanlinna characteristic provides a natural measure of the complexity of a max-plus meromorphic function.

We have shown that many ultra-discrete equations admit infinite-order max-plus meromorphic solutions but the ultra-discrete Painlev\'e equations appear to admit finite-order max-plus meromorphic solutions.  The general solutions of both difference equations and ultra-discrete equations contain arbitrary period one functions.  One significant difference, however, is that many max-plus meromorphic period one functions have infinite order in the complex setting, while all non-constant max-plus meromorphic periodic functions have order two.  

The Painlev\'e property is closely related to the integrability of differential equations.  Integrability is a notoriously difficult concept to define precisely, especially in the context of the differential and discrete Painlev\'e equations.  In the ultra-discrete setting we have already seen that for $|n|>2$, the equation
(\ref{nomax}) admits only infinite-order non-constant max-plus meromorphic solutions.  This is despite the fact that equation (\ref{nomax}) is linear on $(\mathbb{R},+,\times)$, although it is nonlinear as an equation on the max-plus algebra.  Nonetheless, equations of the form (\ref{second-order}) with the degree of $R$ as a max-plus rational function of $y$ do not appear to arise as standard ultra-discretizations of discrete Painlev\'e equations.  For this reason we prefer to think of the finite order condition as an analogue of the Painlev\'e property rather than integrability.

\vskip 10mm

\noindent{\bf \large Acknowledgements}
\vskip 3mm
\noindent
{The first author's research is supported by an EPSRC Advanced Research Fellowship and a project grant from the Leverhulme Trust.  We also acknowledge the support of the European Commission's Framework 6 ENIGMA Network and the
European Science Foundation's MISGAM Network.}

\end{document}